\documentclass{elsart}

\begin{document}

\begin{frontmatter}

\title{Some Considerations about the Bouchaud-Cates-Ravi-Edwards model
  for Granular Flow}

\author{Roberto C. Alamino\thanksref{10} and Carmen P. C. Prado\thanksref{11}}
\thanks[10]{e-mail: alamino@if.usp.br}
\thanks[11]{e-mail: prado@if.usp.br}

\address{Departamento de F\'{\i}sica Geral,Instituto de F\'{\i}sica\\
Universidade de S\~ao Paulo\\
Caixa Postal 66318\\
053015-970, S\~ao Paulo, SP, Brazil}

\begin{abstract}
In this paper we discuss some features of the BCRE model. We show
that this model can be understood as a mapping from a
two-dimensional to a one-dimensional problem, if some conditions are
satisfied. We propose some modifications that 
(a) guarantee mass conservation in the model (what is not assured in
its original form) and (b) correct undesired
behaviors that appear when there are irregularities in the surface of
the static phase. 
We also show that a similar model can be deduced
both from the
principle of mass conservation (first equation) and a simple
thermodynamic  model
 (from which the exchange
equation can be obtained). Finally, we solve the model 
numerically, using different velocity profiles and studying the
influence of the different parameters present in this model.

\end{abstract}


\end{frontmatter}

\section{Introduction}

Since the end of 1980s, when Bak \textit{et al. }used a sandpile as a
paradigm for self-organized criticality\cite{Bak}, the interest in granular
materials experimented a revival. The topic, however, is not recent. The
first studies with granular materials date back to 1773, when Coulomb first
observed that this kind of matter could stand in equilibrium in piles at
certain specific angles and, after that, other famous physicists studied the
topic. Faraday discovered the convective instability in vibrating grains and
Reynolds introduced the notion of dilatancy, just to cite some examples.

The study of granular materials is important since their applications in
industry is wide enough to cover areas as distinct as civil construction and
food transportation. The study of flows can uncover the behavior of dunes,
sand-storms and avalanches. But those are just a few examples. Granular
materials are present everywhere in nature, and in several branches of
industry. Chemical industries, pharmaceutics, mining, geology are just some
more examples of other areas where the study of grains can lead to
important results.

In 1994, a paper by Bouchaud, Cates, Ravi and Edwards \cite{Bouchaud}
presented an important model that became known as the BCRE model. This 
model was successful in describing the qualitative behavior of
flowing grains, and had the additional advantage of being very simple. It
assumes that a two-dimensional sandpile (figure 1), with rolling grains on
its surface, can be divided in two ``phases'' (a static phase $h$ and a
rolling phase $R$) and propose two coupled partial differential equations to
model their behavior:

\begin{equation}
\frac{\partial R \! \left(x,t\right)}{\partial t}=-\frac \partial {\partial
x}\left[ vR(x,t)\right] +\frac \partial {\partial x}\left[ D\frac{\partial
 R(x,t)}{\partial x}\right] +\Gamma  \! \left(R,\tilde{h}\right) \mbox{,}
\label{B1}
\end{equation}

\begin{equation}
\Gamma \left( R,\tilde{h}\right) =-R\left( x,t\right) \left[ \gamma 
\frac{\partial \tilde{h}\left( x,t\right) }{\partial x}+\kappa \,
\frac{\partial ^2\tilde{h}\left( x,t\right) }{\partial x^2}\right] 
=-\frac{\partial \tilde{h}
\left( x,t\right) }{\partial t}\mbox{,}
\label{B2}
\end{equation}

\noindent where $\gamma ,\kappa >0 \,\,$ and 
$\, \tilde{h}=h+x\tan \theta _r, \, \;\; \theta_r$ is the angle of repose.

The first equation defines how the profile of the rolling phase evolves in
time, and the second equation determines the profile of the static phase by
setting the form of the exchange between rolling and static grains,
depending on the local slope of the pile. The model is phenomenological, and
is obtained from educated guesses about the properties observed in real
sandpiles and desired for real grains.

This model gave important clues to how we could describe some interesting
phenomena occurring in granular flow. A variety of papers utilized these
equations to model the behavior of avalanches, stratification and
flows in general\cite{deGennes,deG2,Boutreux,Makse1,Mahadevan}.

The original model, however, is very simplified and  need some
improvements to describe more general situations. Some problems of 
consistency must also
be solved if we want to have a model that do not violate some general
principles, as mass conservation. The model, also, still lacks from a
derivation from first principles or from a microscopic point of view.

In this paper we will address some of the points mentioned above. In section
2, we will show that an equation equivalent to equation (\ref{B1}) can be
obtained from the principle of mass conservation, under the assumption that
the densities of the static and rolling phases are constant in the vertical
coordinate. In section 3, we discuss some limitations of equation
(\ref{B2}) and present some
possible alternatives to it. In section 4, we analyze the consequences of
considering different velocity profiles for the rolling phase and
discuss the
relation of them with other models for granular materials present in
literature, evaluating the role played by some of the parameters 
present in the model.
In section 5, we  present a new and simple model to describe
the mechanism that underline the exchange of grains between static 
and rolling phases. From it we were able to deduce an equation of the
type of equation (\ref{B2}). Finally, in the last section, we summarize
our conclusions.

\section{Mass Conservation}

Consider, for instance, a two-dimensional sandpile with a
rolling and a static phase. The rolling phase is located above the static
phase, and slides over it (see figure 1). Let us assume that the sandpile
can be treated as a continuous medium, $\rho _r$ being the density in area
of the rolling phase. The mass inside an interval $x_o$ to $x_o+\delta x$,
where $\delta x$ is small, is then given by
 
\[
\delta m_r=\int\limits_{x_o}^{x_o+\delta x}\left[ \int\limits_{h\left(
x,t\right) }^{h\left( x,t\right) +r\left( x,t\right) }\rho _r\left(
x,y,t\right) dy\right] dx\mbox{,}
\]

with 

\begin{eqnarray*}
h\left( x,t\right)  &\equiv &\mbox{height of the static phase in a point }x%
\mbox{ at time }t\mbox{,} \\
r\left( x,t\right)  &\equiv &\mbox{height of the rolling phase in a point }x%
\mbox{ at a time }t\mbox{.}
\end{eqnarray*}

We define the linear density of the rolling phase by

\begin{equation}
R\left( x_o,t\right) =\lim_{\delta x\rightarrow 0}\frac{\delta m_r}{\delta x}%
=\lim_{\delta x\rightarrow 0}\frac 1{\delta x}\int\limits_{x_o}^{x_o+\delta
x}\left[ \int\limits_{h\left( x,t\right) }^{h\left( x,t\right) +r\left(
x,t\right) }\rho _r\left( x,y,t\right) dy\right] dx\mbox{.}  \label{defR}
\end{equation}

Remembering that, in general,

\begin{equation}
\lim_{\varepsilon \rightarrow 0}\frac 1\varepsilon
\int\limits_x^{x+\varepsilon }f\left( u\right) du=f\left( x\right) \mbox{,}
\label{limite}
\end{equation}

we can write (\ref{defR}) as 

\begin{equation}
R\left( x_o,t\right) =\int\limits_{h\left( x_o,t\right) }^{h\left(
x_o,t\right) +r\left( x_o,t\right) }\rho _r\left( x_o,y,t\right) dy\mbox{.}
\label{R1}
\end{equation}

When $\rho _r$ is independent of the vertical coordinate $y$, equation
(\ref{R1}) becomes

\begin{equation}
R\left( x,t\right) =\rho _r\left( x,t\right) r\left( x,t\right) \mbox{.}
\label{Rfim}
\end{equation}

Repeating this procedure for the static phase, we obtain 

\begin{equation}
S\left( x_o,t\right) \equiv \lim_{\delta x\rightarrow 0}\frac{\delta m_s}{%
\delta x}=\int\limits_0^{h\left( x_o,t\right) }\rho _s\left( x_o,y,t\right)
dy\mbox{,}  \label{S1}
\end{equation}

and

\begin{equation}
S\left( x,t\right) =\rho _s \!\left( x,t\right)\,  h \! \left(
  x,t\right) \mbox{,}
\label{Sfim}
\end{equation}

where $m_s$ is the mass and $\rho _s$ (again  independent of $y$)
is the density in area of the static phase.

Equations (\ref{Rfim}) and (\ref{Sfim}) define a one-to-one mapping from $h$
and $r$ to $S$ and $R$ respectively, mapping the two-dimensional sandpile in
a one-dimensional problem in the case that the densities of the two
phases do not vary with $y$. A particular case that follows in this
category, used frequently in
literature, is when $\rho_r = \rho_s =$ constant. The possibility of
this mapping  is not really
unexpected, since, with this properties, the variables involved will only depend on the horizontal
coordinate $x$ and time. 

Under the assumption of a continuous model, we can write, for the
rolling phase, a mass conservation equation. For a one-dimensional fluid
of density $R$ that flows under a velocity field $v$ (if  $v$ also 
does not vary in the $y$ coordinate)

\[
\frac{\partial R}{\partial t}+\frac \partial {\partial x}\left( vR\right) =Q%
\mbox{,}
\]

where $Q$ represents the sources or sinks of this fluid. The BCRE
model allows the exchange between rolling and static grains. So, when
describing the rolling phase, the static phase  acts like a
source/sink  at
every point (the extra mass gained by the rolling phase equals the mass
lost by the static phase). We can then write

\begin{equation}
\frac{\partial R}{\partial t}+\frac \partial {\partial x}\left( vR\right) =-%
\frac{\partial S}{\partial t}\mbox{.}  \label{MC}
\end{equation}

Comparing (\ref{B1}) and (\ref{MC}), we can immediately see that the last is
a slightly modified version of the first, where, instead of the height
$h$, we are
working with the density $S$ and the diffusion term (if there is one),
will now depend on the specific shape of the velocity field $v$. 

Equation (\ref{MC}) is  better than
equation (\ref{B1}),  since it guarantees mass conservation for any form of
$v$ (what does not happen with equation (\ref{B1})). If
we assume, for instance, a constant velocity field in (\ref{B1}) (as has
already been done in some previous works [\cite{Makse1,deGennes,deG2}), the
diffusing term must be dropped out in order to ensure mass conservation.

In 1997 Makse \cite{Makse1} applied the BCRE model to a mixture
of two  grains. He wrote equation (\ref{B1}) $-$ with constant velocity
and without diffusion $-$ for each kind of grain

\begin{equation}
\frac{\partial R_i}{\partial t}+v_i\frac{\partial R_i}{\partial x}=\Gamma _i%
\mbox{,\qquad }i=1,2\mbox{,}  \label{Mak}
\end{equation}

and simulated this model with $v_1 = v_2$, showing that, depending on
the value of some parameters, the grains would segregate or stratify.
The assumption $v_1 = v_2$ was not justified in that paper. To
exemplify the advantages of our equations,
applying (\ref{MC}) to a mixture of two grains, and considering that the
mixture is rolling with a constant average velocity $v$, we obtain

\[
\frac{\partial \left( R_1+R_2\right) }{\partial t}+v\frac \partial {\partial
x}\left( R_1+R_2\right) =\Gamma \mbox{,}
\]

that is similar to the equations obtained by Markse (\ref{Mak}).
But now it is easy to see that $v_1 = v_2$ is
not an \textit{ad hoc} assumption, but rather, an imposition of the 
model.

One more advantage of (\ref{MC}) is that now we can obtain a whole series of
different granular flow regimes by varying the velocity profile of the
rolling phase. We will discuss this point better in section 4.

\section{Exchange of grains between static and rolling phases}

We will now make some considerations about the second equation of the BCRE
model, that we call exchange equation. In analogy with (\ref{MC}) we
will first write this equation (\ref{B2}) in terms of 
the new variables $R$ and $S$

\begin{equation}
\frac{\partial S}{\partial t}=R\left[ \gamma \left( \frac{\partial S}{%
\partial x}+\rho _s\tan \theta _r\right) +\kappa \frac{\partial ^2S}{%
\partial x^2}\right] \mbox{.}  
\label{EE}
\end{equation}

In \cite{Boutreux}, Boutreux proposed a  modified version of
this equation, to take into account a shielding
effect, present on the upper grains of the
rolling phase, due to the lower grains of this same layer. Adapted to
the variables $R$ and $S$, this shielding effect can be introduced in
equation (\ref{EE}) as

\begin{equation}
\frac{\partial S}{\partial t}=\frac{R\xi ^{\prime }}{R+\xi ^{\prime }}\left[
\gamma \left( \frac{\partial S}{\partial x}+\rho _s\tan \theta _r\right)
+\kappa \frac{\partial ^2S}{\partial x^2}\right] \mbox{.}  \label{MV2}
\end{equation}

where $\xi ^{\prime }$ is a small constant related to the thickness of the
layer of rolling grains that indeed interact with the static phase. Note
that, if $R\sim \xi ^{\prime }$ (a thin rolling phase), $R\xi ^{\prime
}/(R+\xi ^{\prime })\sim R$, but if $R\gg \xi ^{\prime }$, then $R\xi
^{\prime }/(R+\xi ^{\prime })\sim \xi ^{\prime }$.

Equation (\ref{MV2}) is still not adequate to describe what
happens close to the interface, if there are irregularities on it. 
As the grains flow, they can erode part of the static phase, creating
a (sometimes big) crater with a positive slope in the right border (see
figure 2). To see that, let us analyze a particular case, where the
densities are constant. 

The term $\partial S/\partial x$ is directly related to the slope of
the pile (given by $\partial h/\partial x$). If it  
is negative, the pile is inclined to the
right (and if it is positive the pile is inclined to the left). There
is no problem when the slope is negative. As expected,  for a local
slope above
the repose angle, we have erosion (and for a slope below it we have
acrescion). But, when this term is positive, we always have acrescion,
what is not a reasonable behavior. To correct that we suggest a
modification, where we consider the minus sign of the modulus of $\partial
S/\partial x$

\begin{equation}
\frac{\partial S}{\partial t}=\frac{R\xi ^{\prime }}{R+\xi ^{\prime }}\left[
\gamma \left( \rho _s\tan \theta _r-\left| \frac{\partial S}{\partial x}%
\right| \right) +\kappa \frac{\partial ^2S}{\partial x^2}\right] \mbox{.}
\label{FV}
\end{equation}

\section{Diffusion and  velocity profiles}

One interesting, important and observed property of the BCRE
model is that it includes a diffusion  of the rolling grains. The presence
of diffusion in this phase is observed and, in fact, very
obvious. This diffusion, however, leads to some constraints on the
velocity profile $v$. 

The correct expression for the velocity field $v$, in equation
(\ref{MC}), should, in principle, be derived from a momentum conservation
equation. However, writing such an equation is not an easy task, since it
should take into account all the interactions
between the grains and would depend on the stress tensor of the
material. In general, we can only say that the velocity field is a function
of $x$ and $t$  that depends on a variety of
factors. Most of the works present in literature have assumed a
constant velocity profile, for simplicity.

We will, in this section, analyze two possible functional forms for $v$,
showing some results of computer simulations and
studying its effects in  the shape of the pile. 

In all 
computer simulations we integrate the BCRE equations numerically by means
of a finite difference scheme (see appendix) and
constructed an online animation of the profile of both phases in real
time. The figures
presented bellow are screenshots of the animation generated by the
program. We considered the following profiles:

{\bf(1)} $v = v(R) = \alpha \partial _xR+\beta R+\delta \,$, where $\,
\alpha ,\beta
,\delta$ are constants.

Here $v$ is considered as being a function of  $R$ only,  and  a
kind of gradient expansion was done. Note that, in  this case, 
$v$ is an implicit function of $x$ and $t$ ($v = v(R)$ where $R = R(x,t)$).
Substituting in (\ref{MC}) we obtain

\begin{equation}
\frac{\partial R}{\partial t}+\frac \partial {\partial x}\left( vR\right) =%
\frac{\partial R}{\partial t}+\alpha \left( \frac{\partial R}{\partial x}%
\right)^{\!2}+\left( 2\beta R+\delta \right) \frac{\partial R}{\partial x}%
+\alpha R\frac{\partial ^2R}{\partial x^2}\mbox{.}
\label{IS1} 
\end{equation}

\noindent Note that this functional form of $v$ includes a diffusion term.
 It can also be seen from (\ref{IS1}) that:

i) The diffusion coefficient is proportional to $R$. This means that
the higher the pile, the more it will diffuse. This is reasonable, and
can be understood as a consequence of gravity.

ii) Equation (\ref{IS1}) presents a non-linear term with a constant 
coefficient $\alpha$, ($\alpha \neq 0$). When diffusion is small 
and can be neglected (as in most
cases studied in literature), the non-linear term is also unimportant,
and the equation above is in agreement with previous works.
Diffusion, however, affects the profile of the pile.
Figure 3 shows the effects of this term.  It compares the profile of
a pile of grains, at the same time and for the same initial condition,
for:  (a) the original model with constant
velocity and (b), (c) equation (\ref{IS1}) with $v = \alpha \partial_x R$, 
for two different values of $\alpha$. We can see that
the final shape of the bump is quite different if $\alpha$ is not too small.

iii) This equation  has also an advective term, i.e., a term on the
first derivative of $R$ with
respect to $x$. Its  coefficient depends linearly on $R$ (for $\beta
\neq 0$) \footnote{An advective term like that had already been proposed in 
other papers,
as in \cite{Aradian}}. Figure 4 shows how it affects the profile of the
pile. We can
see that there is a tendency to the formation of
shock fronts in one of the sides of the bump.

{\bf(2)} $v=f\left( \partial _xh\right)$

In a real pile of grains, there may be irregularities with 
positive slope, due to erosion. In this case, the velocity must 
depend on the sign of the slope, or
else we will have avalanches climbing up the pile at the points with
positive slope, with the same velocity as in the negative slope side.
 To correct this defect, we considered the following form for $
v$

\begin{equation}
v=\left\{ 
\begin{array}{ll}
\alpha \, \partial _xR+\beta R+\delta ,\mbox{\qquad } & \mbox{if }\partial
_xh\leq 0 \\ 
v_s, & \mbox{if }\partial _xh>0
\end{array}
\right. \mbox{,} 
\label{velo}
\end{equation}

where $v_s$ is a constant.

For $\partial_x h \leq 0$ this expression is equivalent to the
previous form of $v$. But, for $\partial_x h \ge 0$, the rolling
grains, now, meet a barrier of static grains, and move up this barrier
with a constant
velocity. There is no diffusion in this case, since
the velocity is constant. But now this is a desired property:
the grains slowly accumulate in the barrier (and do not diffuse).
Figure 5 shows the profile of the pile at the same time when (a)
the velocity is independent of the slope and (b), (c) $v$ depends on
the slope according to (\ref{velo}) for two different values of
$v_s$. The static phase is gray, and has been settled to an irregular
shape to amplify the consequences of the modifications introduced.

When we were at the end of this work, we came to know about a work of
Herrmann and Sauermann, on the behavior of the Barchan Dunes
\cite{Herrmann}, that was also based on the BCRE model, and dealt with some of
the points we present in this paper. They present a
two-dimensional version of (\ref{MC}) but they do not deduce it.
In particular,  there is no mention of the need of
independence of the densities with the $y$ coordinate. Indeed, we believe that,
(although it is not explicitly written)
the simulations were performed under constant densities which
is a particular case of $y$ independence. They also used a slightly
different version of (\ref{B2}), where the modulus of the slope of the
static phase were also employed.

\section{Simple Model for the exchange of grains between rolling and
  static phases}

It is possible, on the bases of a naive model, to deduce an equation
to describe the exchange of grains between rolling and static phases.
Remember that the densities of the rolling and static phases are
different, the latter being higher than the former. As the rolling phase
rolls over the static phase, the friction between them takes energy
out of the
rolling grains. Part of this energy becomes heat, but the rest of it
is transferred to the grains of the static phase. This energy will
agitate these grains and the density of the static phase right below the
interface will increase until it attains the critical dilatancy and starts to
move, becoming part of the rolling phase. This process is similar to
a solid to liquid first order phase transition, the static phase 
playing the role of the solid (receives energy and 
starts to ''melt'').  Indeed, there are
experimental evidences that, at least in the case when the transition is
induced by  tilting, it does have features of a first order phase
transition \cite{Jaeger}. We will assume that this analogy is
valid. We can than calculate the mass of static grains
that will "melt'' (that is, receives energy and starts to roll)
using an analogy with the latent heat equation

\[
\delta Q = L \, \delta m_s\mbox{,}
\]

where $\delta Q$ is the energy acquired from the rolling phase, $\delta m_s$
is the mass of static grains that melts and $L$
is a constant, analogue to the latent heat.

Let us now focus on what happens in a little interval $\delta x$ of
the horizontal coordinate of the pile (see figure 1).
The mass that is melted is a portion of the total mass of static
grains. However, not all the static grains receive energy
from the rolling phase. The upper grains shield the lower grains from
the contact with the rolling phase. We will consider that the mass that
can actually receive energy (and, therefore, melt) is a fraction $\xi h$
of the static phase (we will justify this assumption, now adopted for
the sake of simplicity, ahead).

So, the mass that can be melted is given by 

\[
m_s = \rho_s V =  \xi \, h \, \delta  x \,
\rho _s = \xi \, h \,
\delta  x \,  \frac Sh\mbox{,}
\]

where equation (\ref{Sfim}) were used. Then, 
$\delta m_s = \delta  \! S \: \xi \, \delta  x$ and

\begin{equation}
\delta Q =  L \, \xi \: \delta S \, \delta x
\mbox{,}  \label{HS}
\end{equation}

where $\delta S$ is the variation in $S$ caused by the melting of the static
phase. Note that when the pile is too
high, internal forces and gravity act to de-estabilize it, increasing
the melting. This can justify the assumption, made above, that $m_s$
is proportional to $h$ (instead of being a constant layer).

If all the energy to melt the static phase
comes from the rolling phase, and if it is a fraction of the kinetic energy
lost due to friction in the interface, we have

\begin{equation}
\delta Q=c \, \delta K\mbox{,}  \label{BEN}
\end{equation}

where $\delta K$ is the kinetic energy lost by the rolling phase. 
But $K=p^{2}/2m_r$, where $m_r$ is the mass of the rolling phase and $p$
its momentum. So we have

\begin{equation}
\delta K=\frac{2p \, \delta p}{2m_r}=\frac{\left( m_rv\right) \delta
  p}{m_r}%
= v \, \delta p\mbox{.}  \label{KEL}
\end{equation}

Remembering that $m_r = R \, \delta x$, supposing that the rolling
phase transfers energy to the static
phase only by friction, and that the friction is proportional to the
weight of the rolling phase at $x$, we have

\begin{equation}
\frac{dp}{dt}=\mu \,m_r \,g\Rightarrow \delta p=\mu \,m_r\,g \,\delta 
t=\mu \,g \, R \, \delta x \, \delta t\mbox{,}  \label{VOM}
\end{equation}

and, from (\ref{HS}), (\ref{BEN}), (\ref{KEL}) and (\ref{VOM}), we have

\[
L\ \, \xi \, \delta \! S \, \delta x = c \,\mu \,g \,
 v \, R \, \delta t \, \delta x\mbox{.} 
\]

That, in the limit $\delta t$, $\delta S \rightarrow 0$, gives

\begin{equation}
\frac{\partial S}{\partial t}=\beta \,v \,R\mbox{,}  \label{EE2}
\end{equation}

where $\beta = \displaystyle\frac{c \, \mu \, g}{L \,\xi }$ is a constant.

This is a quite simple expression for the exchange equation. It  can assume
a variety of forms, depending on the velocity field $v$. It incorporates
the velocity field explicitly, indicating that the exchange of grains
between both phases  depends on the exact
shape of $v$, what is very reasonable, since the velocity  of the
rolling grains interfere directly with the energy lost in the
collisions, that, ultimately, are responsible for the transformation 
of static grains in rolling grains.

Note also that if (\ref{EE2}) is substituted in the mass
conservation equation, we get
\[
\frac{\partial R}{\partial t}+v\frac{\partial R}{\partial x}=q\mbox{,}
\]
where
\[
q=\left( \beta v-\frac{\partial v}{\partial x}\right) R\mbox{.}
\]

When the velocity is a function of $x$, $t$ and $R$ only,
$q$ will be, in principle, also a function of these three variables
(because $R$ is a function of $x$ and $t$ only), and the resulting equation
will be a well known quasi-linear partial differential equation of 
first order in $R$,
that can be solved by the method of characteristics given by the simple
system
\[
\frac{dt}1=\frac{dx}{v\left( x,t,R\right) }=\frac{dR}{q\left( x,t,R\right) }%
\mbox{.}
\]

The only difficulty is that, if $v$ has an explicit dependence on $R$%
, the variable $q$ will have a dependence in $\partial R/\partial x$,
what may
make the system of characteristics difficult to be solved analytically
in practice. However, a solution for $R(x,t)$ will give consequently a
solution for $S(x,t)$ by means of (\ref{EE2}). 
We intend to further explore
this point in a following paper.

Furthermore, if equations (\ref{EE2}) and (\ref{FV}) are  equivalent,
the velocity field must assume the form of

\[
v=\frac{\xi ^{\prime }/\beta }{R+\xi ^{\prime }}\left[ \gamma \left( \rho
_s\tan \theta _r-\left| \frac{\partial S}{\partial x}\right| \right) +\kappa 
\frac{\partial ^2S}{\partial x^2}\right] \mbox{.} 
\]

This velocity field  has some interesting features. First,
note that $v$ depends on the slope of the 
pile. The first term inside the square brackets
becomes positive for a slope below the angle of repose and negative above
it, what means  that the velocity is higher where the pile is
steeper. Second, 
the factor $\left( R+\xi ^{\prime }\right) ^{-1}$ suggests that $v$
is inversely proportional to the  weight of the pile of rolling
grains, what is reasonable.

\section{Conclusions}

In the first two sections of this paper, we pointed out some problems
with the equations of the BCRE model, suggesting the following
modification:

\begin{equation}
\frac{\partial R}{\partial t}+\frac \partial {\partial x}\left( vR\right) =-%
\frac{\partial S}{\partial t}\mbox{,}
  \label{FBE1}
\end{equation}

\begin{equation}
\frac{\partial S}{\partial t}=\frac{R\xi ^{\prime }}{R+\xi ^{\prime }}\left[
\gamma \left( \rho _s\tan \theta _r-\left| \frac{\partial S}{\partial x}%
\right| \right) +\kappa \frac{\partial ^2S}{\partial x^2}\right] \mbox{.}
\label{FBE2}
\end{equation}

The first equation explicitly assures  mass conservation.
The second equation was modified to take into
account the signal of the slope in the static phase. In addition, we
introduced the new variables $R$ and $S$, giving
a precision definition for them, a point that 
was not totally clear in the literature until now.

We used the corrected equations (\ref{FBE1}) and (\ref{FBE2}) to simulate two
possible velocity fields, and  found acceptable results.

Finally, we have proposed a model for the exchange of grains between the
rolling and static phases. From it, we  obtained the alternative 
$\: \displaystyle \frac{\partial S}{\partial t} = \beta \, v \, R \:$
for the exchange equation, that is simpler, includes
explicitly the velocity field, and lead to interesting results.

The  model presented is very simple, and we are aware that
many of the hypothesis  made must be examined in more detail.
We assumed that the friction is proportional to the weight of the
rolling phase, and this is surely oversimplified. Probably there is a more 
complicated dependence on other parameters of the model as well. For instance,
it is reasonable to suppose that the energy transfered to the static
phase also  depends on the shape of the grains, its
density, the toughness of the material and a variety of other
factors. 

Also, we think that the fraction of static grains that receives
energy from the rolling phase is probably a more general function of
$h$, not to speak of a dependence on  the other variables of the model.

We have also neglected the inverse process, i.e.,
the transformation of rolling grains in static ones. Our model 
may be good to describe an avalanching process,
where the inertia of the rolling grains is large,
but may fail to describe a more general situation.

We hope to address this points in a following paper. However, we think
it is  amazing and interesting that a somewhat rich expression for $v$, 
like (\ref{EE2}), can be obtained from such a naive model.

Acknowledgements: R. C. A. acknowledges the financial support of the 
Brazilian agency FAPESP.

\section{Appendix}

The equations of the BCRE model were  integrated with the
operator splitting method \cite{split}. Suppose a differential equation
of the form

$$
\frac{\partial u}{\partial t} = L(u),
$$

where $L(u) = \sum_{i=1}^{m}L_i(u)$ is a generic non-linear operator
that can be written as a sum of $m$ other operators, and $u$ is a function of
$x$ and $t$. If we have a good method to integrate each of the equations
$\displaystyle \frac{\partial u}{\partial t} = L_i(u)$, then $u^{n+1}$ 
can be obtained through $m$ successive time steps:

\begin{eqnarray*}
u^{n + (1/m)}\; & = & \;  L_1(u^n, \, \delta t),\\
u^{n + (2/m)} \; & = & \; L_2 \left (u^{n+(1/m)}, \, \delta t \right )\\
\vdots \; \; \; \; \; \; & = & \; \; \; \; \; \; \vdots \\
u^{n+1} \; & = & \; L_m \left (u^{n+(m-1)/m}, \, \delta t \right ).
\end{eqnarray*}

For the first equation of the model (and its variants), the $L_i$
operators are of the form

$$
L_1(r) = f(r)\frac{\partial r}{\partial x}, \quad L_2(r) = g(r)
\frac{\partial^2 r}{\partial x^2}, \quad L_3(r) = q(r) \;
\mbox{and} \quad L_4(r) = k
\left( \frac{\partial r}{\partial x} \right)^{\!2},
$$

\noindent where $f$, $g$ and $q$ are arbitrary functions of $r$ , 
and $k$ is a constant.

$L_1$ and $L_2$ where integrated with variants of the Crank-Nicholson
method; the operator $L_3$ was integrated with a fourth order
Runge-Kutta procedure and finally $L_4$ was integrated with a FTCS
finite-difference scheme.

The second equation can be split into two operators of the form:

$$
L_1 \left(h \right) = a \frac{\partial^2 h}{\partial x^2},
\qquad L_2 \left( h \right) = b \frac{\partial h}{\partial x},
$$

\noindent where $a$ and $b$ are constants. Now $L_1$ was integrated
with the aid of Crank-Nicholson and the operator $L_2$ with the aid of
a two-step Lax-Wendroff \cite{split} procedure. 

The program was run {\it on line} in a Intel Pentium III 400 Hz or in 
a AMD K6II 400 Hz (about 10 seconds for a single run in both cases).

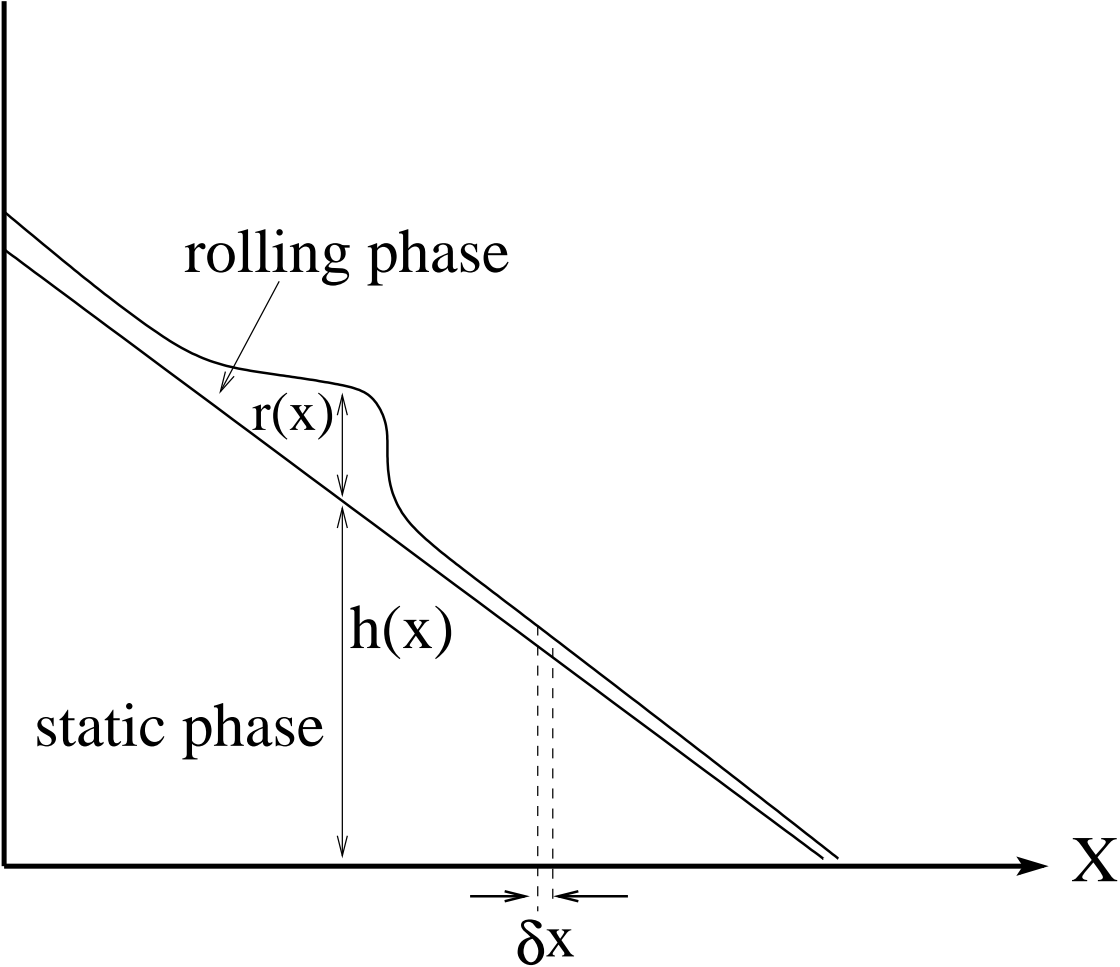
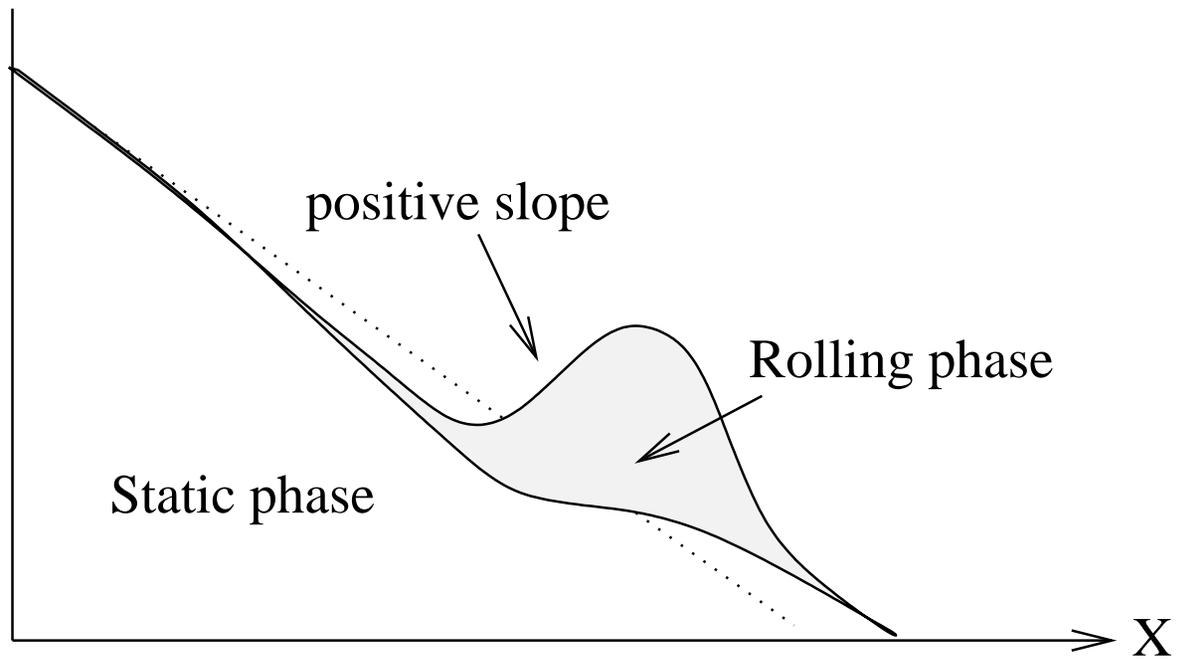
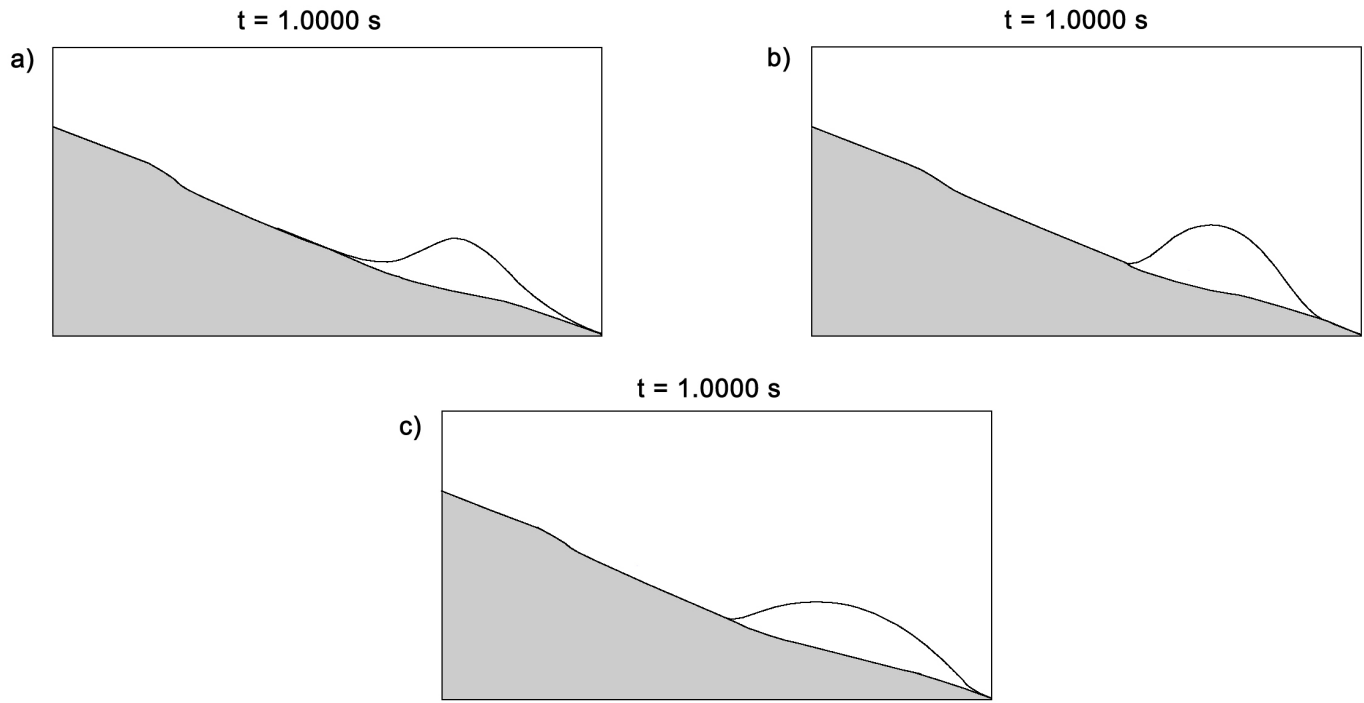
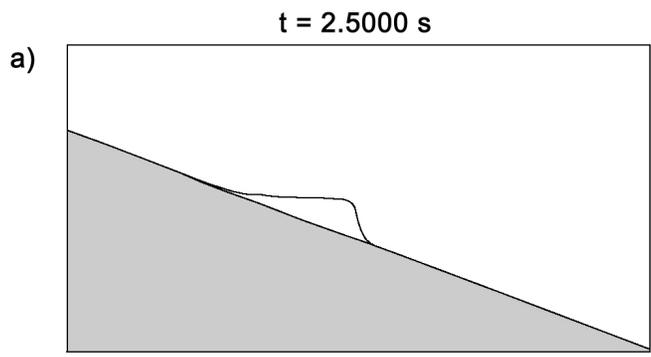
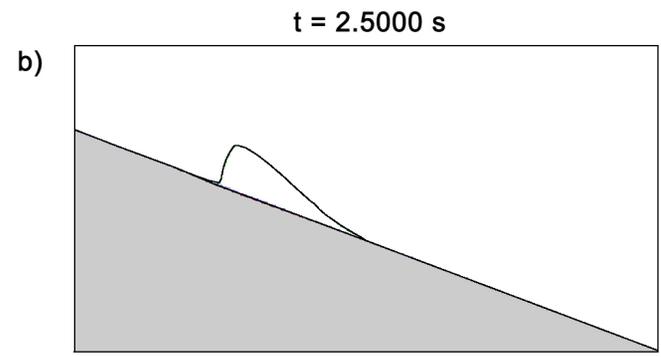
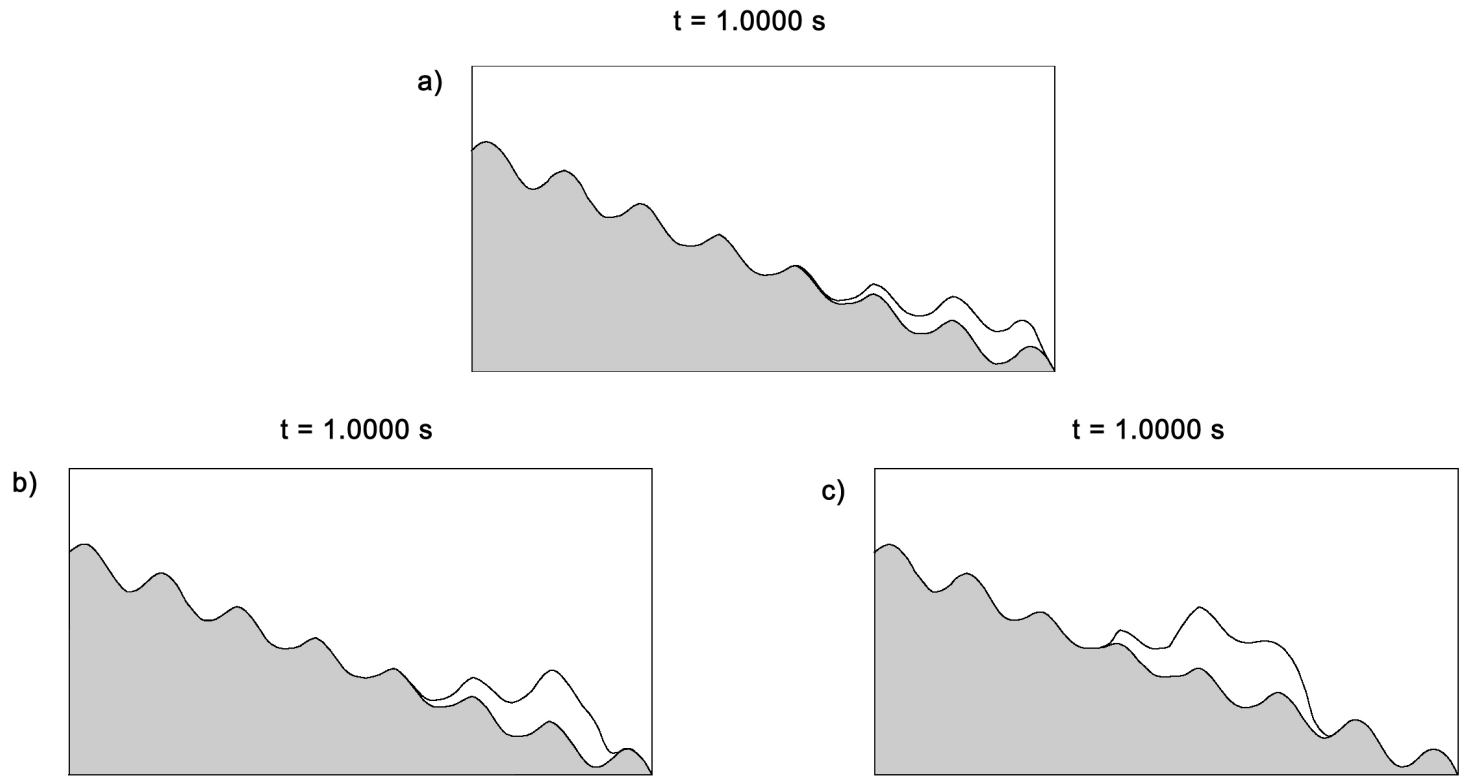
\end{document}